\documentclass{pasj00}
\draft

\begin{document}
\SetRunningHead{I. Matsuyanagi et al.}{Near Infrared Survey of BRC 14}
\Received{2000/12/31}
\Accepted{2001/01/01}

\title{Sequential Formation of Low-Mass Stars in the BRC 14 Region\footnotemark[\ast]}

\footnotetext[$\ast$]{The electronic table (E1) is available at the www site of
(http://pasj.asj.or.jp/xxx)}
%

%
\author{
  Ikuko  \textsc{Matsuyanagi}\altaffilmark{1},
  Yoichi \textsc{Itoh}\altaffilmark{1},
  Koji \textsc{Sugitani}\altaffilmark{2},
  Yumiko \textsc{Oasa}\altaffilmark{1},
  Tadashi \textsc{Mukai}\altaffilmark{1}, \\
and
  Motohide \textsc{Tamura}\altaffilmark{3}}
\altaffiltext{1}{Graduate School of Science and Technology, Kobe University, 1-1 Rokkodai, Nada, Kobe, Hyogo 657-8501}
\email{yitoh@kobe-u.ac.jp}
 \altaffiltext{2}{Graduate School of Natural Sciences, Nagoya City University, Mizuho-ku, Nagoya 467-8501}
\altaffiltext{3}{National Astronomical Observatory, 2-21-1 Osawa, Mitaka, Tokyo 181-8588}

\maketitle

\begin{abstract}
We have carried out a deep near-infrared survey of a bright-rimmed 
molecular cloud, BRC 14 (IC 1848A). 
The 10$\sigma$ limiting magnitude of the survey is 17.7 mag 
at the $K$-band. 
Seventy-four sources are classified as young stellar object (YSO) candidates 
based on the near-infrared color-color diagram. 
The faintest YSO candidates may have masses of 
an order of tenths of the solar mass,
assuming the age of 1 Myr.
We examined three values as indicators of star formation;
fraction of the YSO candidates, extinctions of all sources, 
and near-infrared excesses of the YSO candidates.
All indicators increase from outside of the rim to the center of the 
molecular cloud, which
suggests that the formation of the low-mass stars in the
BRC 14 region proceeds
from outside to the center of the cloud.
\end{abstract}

%
%
%
  
\section{Introduction}

Recent analyses of meteorites confirm the presence
of Fe isotopes in the early solar nebula (Tachibana \&  Huss 2003).
This fact suggests
that the Sun was formed in a high-mass star forming region that
has expericenced at least one supernova explosions
(Hester et al. 2004).  
It is important to understand
the formation process of low-mass stars in environments under the influence
of massive stars.

Bright-rimmed clouds (BRCs) are considered to be such sites located in
the peripheries of H\emissiontype{II} regions excited by massive stars.
Sugitani, Tamura, Ogura (1995) obtained near-infrared (NIR) images of
44 bright-rimmed clouds which harbor IRAS point sources in the SFO catalog
(Sugitani, Fukui, Ogura 1991). They
found that small NIR clusters are situated from the IRAS source toward
the tips of some BRCs but mostly behind their bright rims.  This asymmetric
distribution of the clusters strongly suggests star formation propagation from
the side of the H\emissiontype{II} region toward the IRAS source due 
to UV from massive
stars, i.e., small-scale sequential star formation.  This idea was further
supported by a grism spectroscopic survey of H$\alpha$ emission stars
of BRCs (Ogura, Sugitani, Pickles 2002).

However, the limiting magnitude of Sugitani et al. (1995)  was not 
deep enough to
detect sub-solar mass young stellar objects (YSOs).
In fact, the detected sources are considered to be mainly YSOs of 
$\sim$1-2 \MO
based on their $K$-band magnitude.
It is necessary to make deeper NIR observations in order to fully reveal
the cluster members associated with the BRCs and to examine the details of
star formation there.

BRC 14 (IC 1848A) is considered to be one prominent
example of such sequential star formation.
The bright rim of BRC 14 is an ionization front at the interface
between the hot ionized gas of the H\emissiontype{II} region IC 1848 (S 199)
and the cold dense material of the molecular cloud, where
a bright infrared source AFGL 4029 (IRAS 02575+6017) is embedded.
BRC 14 is a part of active high-mass star
forming region IC 1848, whose age is estimated to be $\leqq$
1 Myrs (Harris 1976; Feinstein et al. 1986; Deharveng et al. 1997).
Its photometric distance is 2.2 $\pm$ 0.2 kpc
(Becker \& Fenkart 1971; Moffat 1972; Deharveng et al. 1997).

Here, we present the results of the deep NIR observations of BRC 14.
Based on the photometry of NIR sources including those of sub-solar masses,
we will show clear evidence for the sequential star formation of this 
cloud located
in the periphery of IC 1848.

\section{Observations \& Data Reduction}

The observations were carried out in 2001 August 26 with the 
Simultaneous-3 color InfraRed Imager for Unbiased Survey (SIRIUS) 
mounted on the 2.2 m telescope of University of Hawaii. 
SIRIUS takes $JHKs$ images simultaneously
using three 1024 $\times$ 1024 
arrays with a pixel scale of $\timeform{0.288''}$.
Details of the camera are described in Nagashima et al. (1999) and
Nagayama et al. (2003).
One quadrant of the $J$-band array was not functional at the time of this
observation.
We imaged a $\timeform{4.9'}\times\timeform{4.9'}$ area 
centered on $\alpha=\timeform{03h01m31s}, \delta=\timeform{+60D28'52''}$ 
(J2000), and the same size field with $\timeform{-100''}$ offset in RA,
in order to compensate for failure of the $J$-band array.
We obtained 30 dithered images per each area with an exposure time 
of 30 s for each frame.
A part (9.8 arcmin$^{2}$) has twice as much total integration 
time as else.
Sky frames, offset $(\alpha, \delta) = 
(\timeform{-120''}, \timeform{+500''})$
from BRC 14, were taken just before and after the object frames. 
We observed the standard star P9107 (Persson et al. 1998) 
for photometric calibration.
The typical seeing size was measured to be $\timeform{1.0''}$ at $Ks$.
Dome flat frames were obtained at the beginning of the night.

To reduce the data, we used the Imaging Reduction and Analysis Facility 
(IRAF) software.  
First, linearity was corrected for all frames.
After subtraction of the median-combined sky frame, 
each image was divided by the normalized flat frame. 
Then we combined them into one frame for each band.
To identify NIR sources, we used the DAOFIND task in IRAF 
with a 3 $\sigma$ detection threshold above the background. 
The APPHOT package was used for photometry with a $\timeform{3.5''}$
diameter aperture (6 pixel radius).
The annular sky area was chosen with an inner radius of $\timeform{2.9''}$ and 
an outer radius of $\timeform{4.3''}$.
All magnitudes are transformed into the CIT $JHK$ system.
The 10 $\sigma$ limiting magnitudes for the photometry 
are 19.2, 18.5, and 17.7
at the $J$, $H$, and $K$ bands, respectively.
Bright sources with $J,H\lesssim12$, or $K\lesssim11$  are saturated.
The coordinates of the objects
are determined by three stars listed in the USNO-B1.0 catalog with
the accuracy of about $\timeform{0.5''}$.

\section{Results}

The $JHKs$ composite image of BRC 14 is shown in figure \ref{image}.
The bright rim of BRC 14 is clearly seen in the NIR image.
The position of the NIR rim is coincident with the rim in the optical
wavelengths.
Nebular emission components of AFGL 4029, 
IRS 1 (west) and IRS 2 (east), are much brighter in the NIR wavelengths
than in the optical wavelengths. 
Many invisible sources are also detected in the NIR wavelengths.
We have detected 607 sources within 10 $\%$ photometric accuracies
in all three bands.

We select YSO candidates, based on their NIR excess 
originated from a circumstellar envelope and/or disk.
An upper-left panel of Figure \ref{col-col} is the ($J-H$, $H-K$) 
color-color diagram for the entire survey region.
We define three regions in the color-color diagram (Itoh et al. 1996).
The ``P'' region is the region between the reddening lines extending from 
the loci of main-sequence stars and giants.
The objects plotted in this region are interpreted as main-sequence stars, 
giants, supergiants, Class III sources, or Class II sources with
small NIR excess.
The ``D'' region, where Class II sources are mainly plotted, 
is sandwiched between the ``P'' region and the reddening line projected
from the point of $(J-H, H-K)=(1.1, 1.0)$.
This point corresponds to the reddest intrinsic color of
classical T Tauri stars (CTTSs, Meyer et al. 1997).
Redward of the ``D'' region is the ``E'' region, in which Class I sources 
are plotted.
We classify YSO candidates with NIR-excess (hereafter NIR-YSO candidates) 
as the objects plotted in the ``D'' or ``E'' region 
only if its 1 $\sigma$ photometric error bars lie entirely within those regions.

With this criterion, 74 sources are identified as NIR-YSO candidates 
(table E1).
The objects are identified between the previous NIR observations 
and our observations,
if differences of the position and magnitudes of the object 
are less than $\timeform{0.5"}$ and 0.5 mag.
No NIR-YSO candidate corresponds to the NIR sources in Deharveng et al. (1997), 
because most of them are plotted
in the ``P'' (field star) region and some bright sources are saturated 
in our observation. 
Ogura et al. (2002) found 44 emission-line stars in our field of view.
Among them, only 10 objects are identified as NIR-YSO candidates by our NIR survey.
One reason for this low frequency
is that weak-line T Tauri stars have the similar NIR color to dwarfs.
26 emission-line stars have weak H$\alpha$ emission line
(W(H$\alpha$)$<$ 10 \AA).
The other reason for low frequency is that even for CTTSs
half of them cannot be distinguished from field stars only with NIR photometry
(Meyer et al. 1997).

\section{Discussion}

%
%
%
%

\begin{figure}
\begin{center}
\FigureFile(180mm,180mm){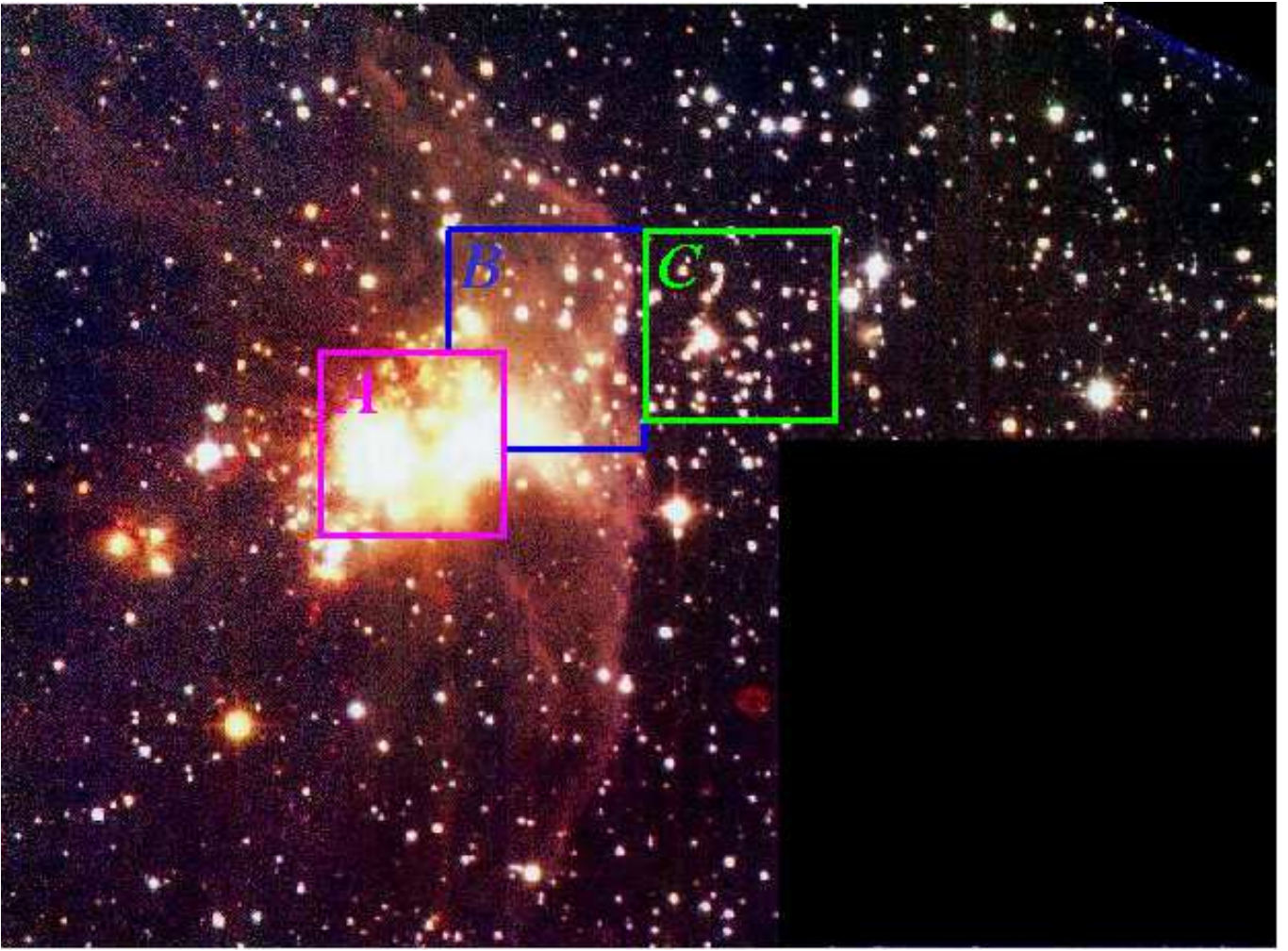}
\caption{$JHKs$ composite image of BRC 14 
($J$ as blue, $H$ as green, and $Ks$ as red). 
The field of view is $\timeform{6.6'}\times\timeform{4.9'}$. 
North is up and east toward the left. 
Two luminous sources are AFGL 4029-IRS 1 (west) and IRS 2 (east). 
The north-south rim is west of AFGL 4029. 
The extended H\emissiontype{II} region S 199 are located west 
(right) of the rim. 
Three areas, A, B, and C are shown for discussion.
Lacks on the top-right corner and the bottom-right square are caused 
by failures of the $J$-band array.}
\label{image}
\end{center}
\end{figure}

\begin{figure}
\begin{center}
\FigureFile(180mm,180mm){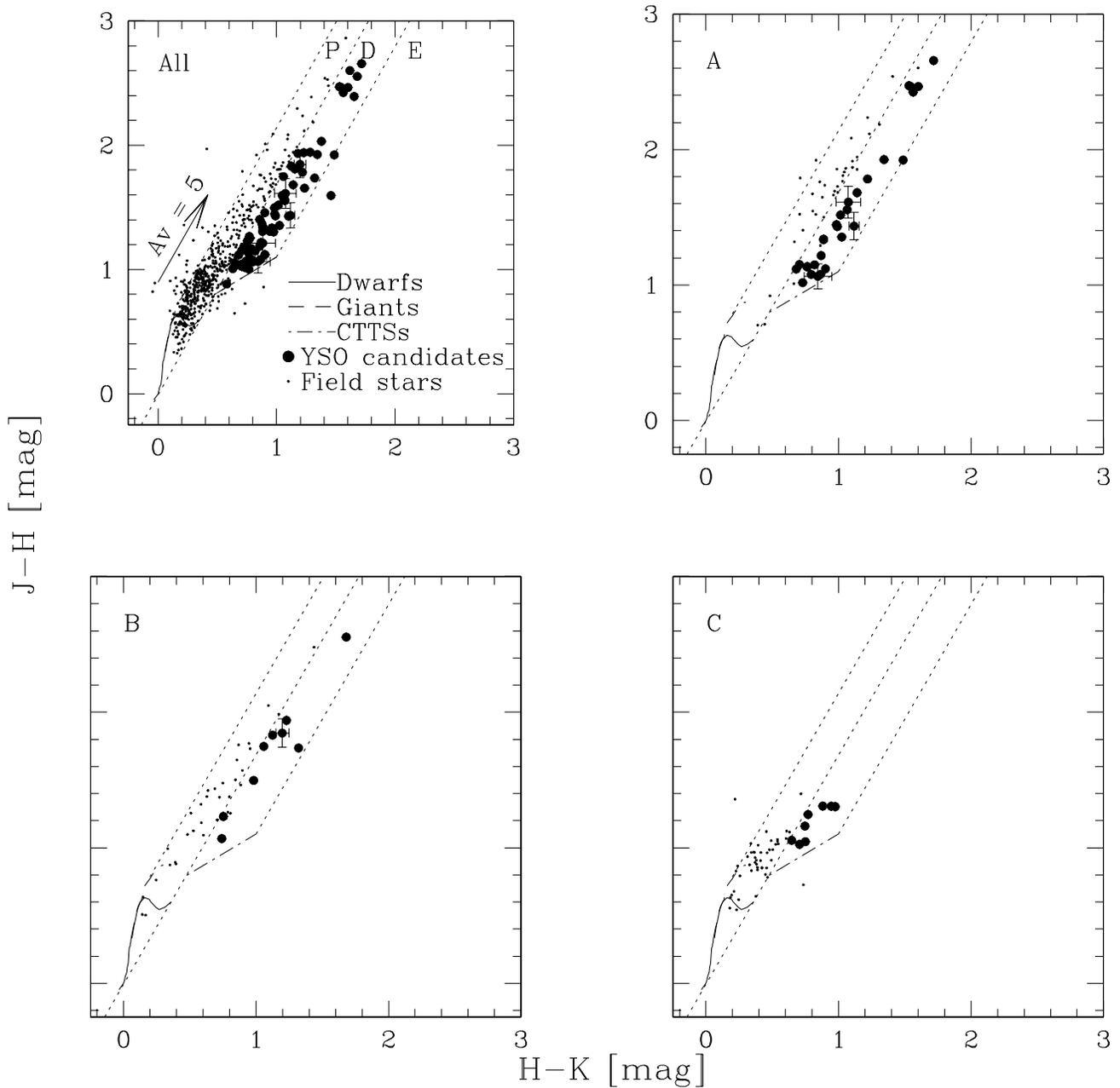}
\caption{Color-color diagrams of the objects in the entire region and
the areas A, B, and C. 
Only the objects with their photometric uncertainties smaller than 0.1 mag 
in the three bands are plotted. 
Error bars are shown in the case of the NIR-YSO candidates with uncertainties
$\geqq$ 0.1 mag in ($J-H$) or ($H-K$).
Plotted lines are the intrinsic colors of main-sequence stars, 
giants (Bessell \& Brett, 1988), and unreddened CTTSs (Meyer et al. 1997). 
A reddening vector is also shown (Meyer et al. 1997). 
All colors are transformed into the CIT system.
In the area A, many sources have large extinctions, 
and there are a number of NIR-YSO candidates. 
Extinction at the area B is similar amount to that in the area A, 
while the NIR-excesses of the YSO candidates are smaller than 
those in the area A. 
In the area C, there are little sources having large extinction, 
and only the small number of the NIR-YSO candidates are detected.}
\label{col-col}
\end{center}
\end{figure}

The H$\alpha$ emission stars are found as an aggregate in the vicinity of
the bright rim with offset toward the exciting stars of S 199
(Ogura et al. 2002).
The alignment of the objects from west to east, 
i.e. the exciting stars, the H$\alpha$ stars, 
the bright rim, and AFGL 4029, implies that star formation proceeds from 
the H\emissiontype{II} region to the AFGL 4029.
In order to conform the propagation of star formation, 
we investigate characteristics of the YSO candidates in three areas. Each area has
about 1 arc-minute square as shown in figure \ref{image}.
The area A centered on ($\alpha$, $\delta$)=($\timeform{03h 01m 33s}$,
$\timeform{+60D 29' 19''}$) contains the luminous source AFGL 4029. 
The area B is in the bright rim and west of the area A. 
The area C is west of the rim. 
Figure \ref{col-col} represents color-color diagrams of the
objects in the three areas.

We discuss three indicators; extinctions of all sources, 
fractions of the NIR-YSO candidates, 
and NIR-excesses of 
the NIR-YSO candidates (table \ref{parameters}).
Average of the extinctions of the sources in the area
represents the column density of the molecular cloud. 
An extinction of each Class II source is derived from the distance 
between the intrinsic color of CTTSs and the observed color 
on the color-color diagram (Meyer et al. 1997).
On the other hand,
it is difficult to estimate the extinctions of the field-stars.
Since we do not know spectral types of field-stars, we can not
distinguish between a late-type dwarf with small extinction and an
early-type dwarf with large extinction.
A star count model of the Galaxy (e.g. Jones et al. 1981) predicts
that the majority of the field-stars detected by such a deep NIR survey 
are late-type dwarfs.
We assume that the intrinsic color of the field-stars is
the line connecting the intrinsic colors of M4V and K7V.

The sources we detected are categorized into NIR-YSOs, YSO without NIR excess,
and field stars.
We estimate the number of the field stars in each area.
First we count the 2MASS sources down to $K < 14$ mag in 
the region $(l,b)=(\timeform{138D.3}, \timeform{-1D.56})$, 
the opposite side of the Galactic plane.
Then, we calculate the number of the field stars using a star count model of
the Galaxy (Jones et al. 1987). 
The number of the field star is overestimated, because
the model is valid only for high galactic latitute.
The number is well reproduced (difference $<$ 5 \%),
if we take an additional extinction of $A_{\rm K}=0.2$ mag.
By extrapolating this model and taking average extinction
in each area, we estimate the number of field star to be
15, 18, and 24 in the areas A, B, and C, respectively.
The numbers of YSOs are then estimated to be 50, 24, and 32 in the
areas A, B, and C, respectively.
Fraction of NIR-YSO candidates is a ratio of the number of NIR-YSO candidates
to the estimated number of YSOs (see table 1).
Haisch et al. (2001) analyzed infrared ($JHKL$)
colors for several young clusters 
then obtained the relation that the fraction of NIR-YSOs
to all YSOs decreases with increasing the age of the cluster.

NIR-excess index of each YSO candidate is also derived from the color-color diagram. 
We define the index as the distance from the color of the object to 
the boundary between the ``P'' and ``D'' regions (the PD boundary), 
which is then normalized by the distance between the PD boundary and 
the boundary between the ``D'' and ``E'' regions (the DE boundary).
An object with the NIR-excess of 0 is plotted on the PD boundary, 
while that with the NIR-excess of 1 is plotted on the DE boundary.  
The NIR-excess is a function of many parameters, such as
accretion rate, an inner radius of a disk, 
and an inclination angle (Hillenbrand et al. 1998).
Assuming random distribution in inclination,
we consider that accretion rate and an inner radius of a disk 
mainly influence on the NIR-excess. 
As discussed in Oasa et al. (2006), younger YSOs have larger NIR-excesses.

If a cluster is as young as the associated YSOs which have optically-thick
disks and are heavily embedded in the parent molecular cloud, 
all indicators above are expected to be large.
In contrast, if a cluster is as old as the YSOs which do not
have optically-thick circumstellar disks and are no longer heavily
embedded, all indicators should be small.
In BRC 14, all indicators decrease in order of the areas A, B, and C.
Many sources in the area A have large extinctions, 
and the area A contains many NIR-YSO candidates.
In the area B, many sources also have large extinctions, 
but the fraction of the NIR-YSO candidates as well as the
NIR-excesses of the YSO candidates are
smaller than those in the area A.
In the area C, the fraction of the NIR-YSO candidates is small, 
and the sources with large extinctions are less abundant compared to 
the areas A and B.
Although these indicators have large uncertainties, 
all indicators are large for the area A and small for the area C, 
implying that the area A is young and the area C is relatively old.
We emphasize that all indicators of the area A are significantly larger 
than those of the areas B and C.

\begin{figure}
\begin{center}
\FigureFile(100mm,100mm){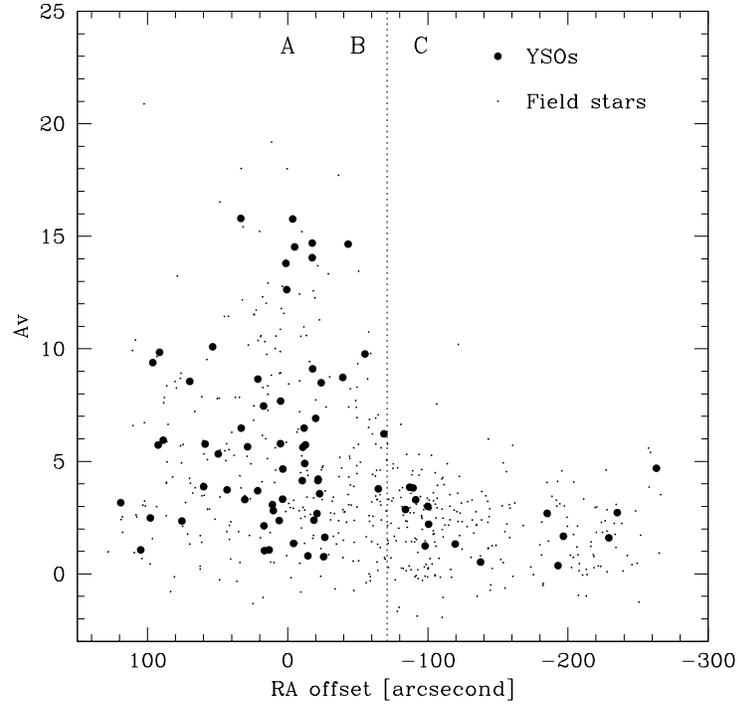}
\FigureFile(100mm,100mm){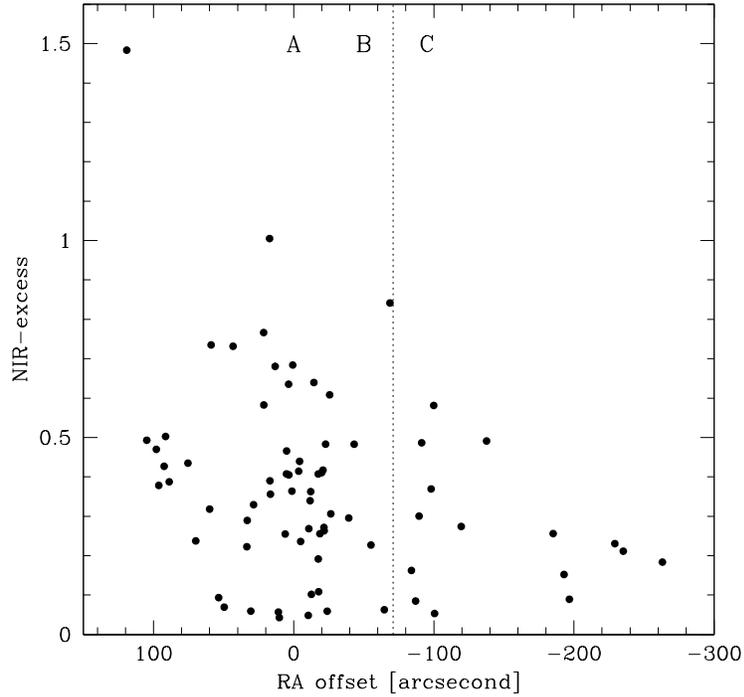}
\caption{Visual extinctions of all sources and NIR-excesses 
of the YSO candidates 
as functions of R.A.. R.A. offsets to the center of the area A 
($\alpha=\timeform{03h 01m 33s}$). 
The dotted lines are the boundary between the areas B and C, 
representing to the position of the rim. 
The left region of the boundary corresponds to the direction toward
the molecular cloud.}
\label{RA}
\end{center}
\end{figure}

The extinctions of all sources and the NIR-excesses of 
the YSO candidates are also shown as functions of
the right ascension of the objects (figure \ref{RA}).
We notice that the extinctions of the sources and 
the NIR excesses of the YSO candidates increase toward the east. 
The extinctions change significantly at the rim.
The figures again indicate that the YSOs with circumstellar structures
are associated with the molecular cloud.
We claim that the sequential star formation previously proposed by the
optical study (Ogura et al. 2002) also occurs inside of the rim
where NIR light can penetrate.

\begin{figure}
\begin{center}
\FigureFile(180mm,180mm){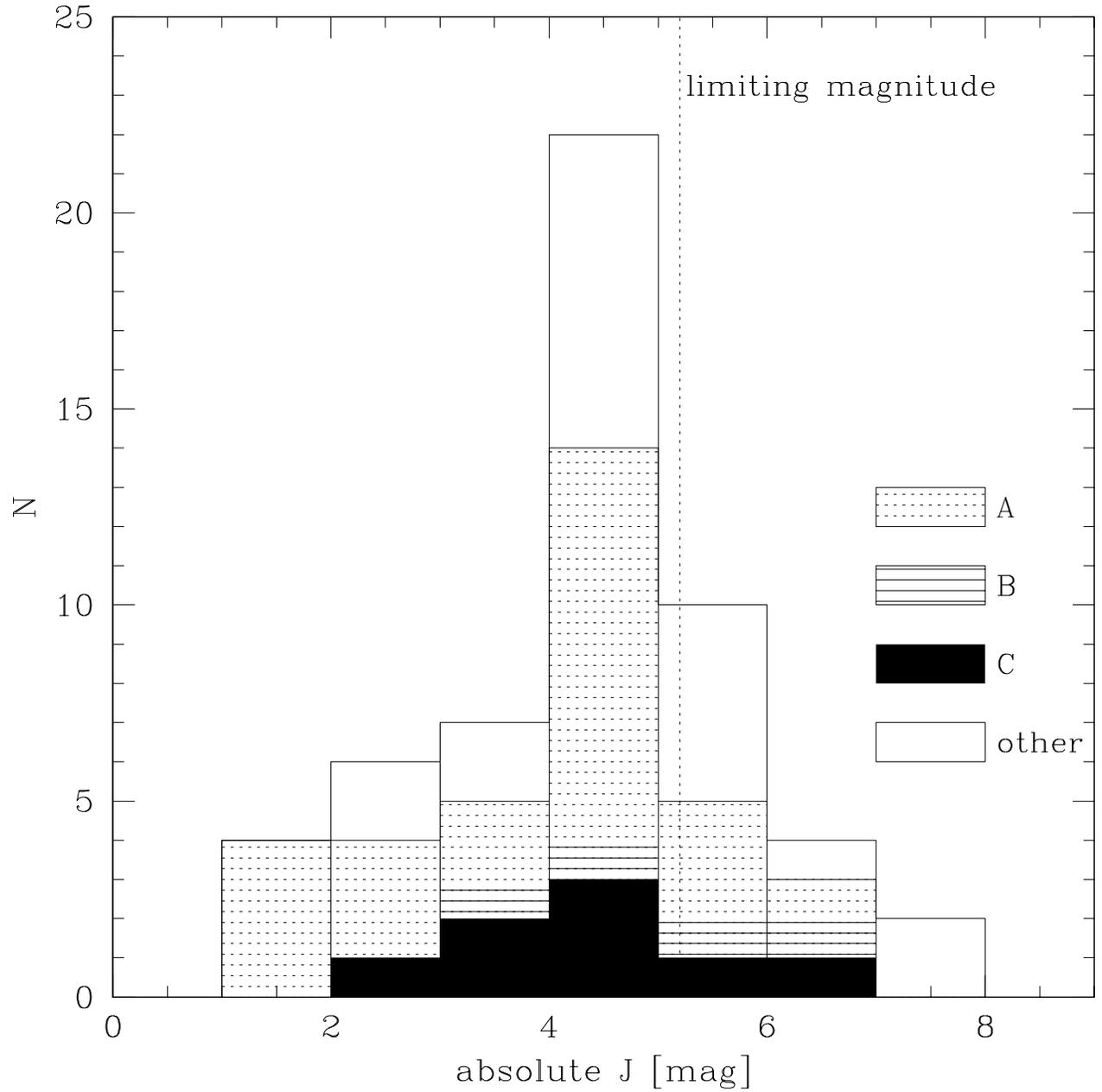}
\caption{
Extinction-corrected $J$-band luminosity function of the
Class II candidates. The sample is limited to the candidates
with $A_{\rm V} \leqq 8$ mag. 
The limiting magnitude of the survey is 5.2 mag in the extinction-corrected
absolute $J$-band magnitude,
corresponding to 0.12\MO.
Low-luminosity YSO candidates are found in all of the three areas.}
\label{lf}
\end{center}
\end{figure}

In all areas, low-mass objects are formed.
To make a census of low-mass YSOs in BRC 14, 
we construct an extinction-corrected
NIR luminosity function of the Class II candidates
plotted in the ``D'' region in the color-color diagram (figure \ref{lf}).
Note that heavily embedded YSO candidates cannnot be detected
even if their intrinsic luminosity is brighter 
than the limiting magnitude.
To remove this bias of extinction, we discuss $J$-band luminosity function 
only for the Class II sources with $A_{\rm V} \leqq 8$ mag.
As $A_{\rm J} = 0.288 A_{\rm V}$ (Bessell \& Brett 1988), $A_{\rm V} = 8$ mag
corresponds to $A_{\rm J} = 2.3$ mag.
The apparent $J$-band limiting magnitude is then 16.9 mag, when
we consider the extinction bias.
With the distance to BRC 14 of 2.2 kpc,
the limiting magnitude in the $J$-band luminosity
function is about 5.2 mag in absolute
magnitude.
By the 1 Myr isochrone of the evolutionary track of low-mass
objects (Baraffe et al. 1998),
the limiting magnitude corresponds to 0.12\MO.
Figure \ref{lf} indicates that
luminous YSO candidates are located only in the area A. 
They might be younger or massive sources.
On the other hand, low-luminosity YSO candidates are found in all of the 
three areas.

An aggregate of NIR sources has been discovered by
Sugitani et al. (1995). 
Its asymmetric distribution leads them to an idea of
sequential star formation in the BRC 14 region, 
though only qualitative discussion was presented.
Ogura et al. (2002) found many H$\alpha$ emission stars in the
BRC 14 region. Most of them are located outside of the rim, 
i.e. in the H II region between the exciting star and the rim. They
proposed small scall sequential star formation from the 
exciting star to the rim.
However, only a small number of the emission-line stars were detected inside
the rim due to heavy extinction of the cloud.
We detect the aggregate of NIR sources inside the rim.
By applying the YSO indicators as above,
we firstly find quantitative evidence that
low-mass stars,
including sub-solar mass ones, in BRC 14 
sequentially form from the outside of the rim to the inside of the molecular
cloud.

%
%
%

\bigskip
\bigskip


\bigskip

We are grateful to S. Sato, T. Nagata, A. J. Pickles, Y. Nakajima,
H. Nakaya, T. Nagayama, C. Nagashima, for the observations.
This work is supported by "The 21st Century COE program: The Origin and
Evolution of Planetary Systems" of the Ministry of Education, Culture,
Sports, Science, and Technology (MEXT). Y.I. is supported by a
Grant-in-Aid for Scientific Research No. 16740256.

\clearpage

\begin{table}[h]
\begin{center}
\caption{Star formation indicators for the three areas.}
\label{parameters}
\begin{tabular}{cccc}
\hline\hline
area       & Av [mag]\footnotemark[1] & NIR-YSO fraction \footnotemark[2]      &  NIR-excess\footnotemark[3] \\
\hline
A      & 8.4 (4.8)    & 29/65 - 15 (58\%) &  0.41 (0.20)\\
B      & 6.9 (4.0)     & 9/42 - 18 (38\%)&  0.29 (0.23)\\
C      & 3.0 (1.5)     & 8/56 - 24 (25\%)  &  0.29 (0.17)\\
\hline
\end{tabular}
\end{center}

\footnotemark 
Average of the extinctions of all sources and its standard deviation in parenthesis.

\footnotemark   
(the number of the YSO candidates with NIR-excess)/(the number of all sources - the expected number of field stars). Its percentage is shown in parenthesis.

\footnotemark    
Near-infrared excess of the YSO candidates and its standard deviation in parenthesis.
\end{table}


\begin{thebibliography}{}
\bibitem[Baraffe et al.(1998)]{Baraffe}
  Baraffe, I., Chabrier, G., Allard, F., Hauschildt, P. H. \ 1998, \aap, 337, 403
\bibitem[Becker et al.(1971)]{Becker}
  Becker, W., Fenkart, R. \ 1971, \aaps, 4, 241
\bibitem[Bessell et al.(1988)]{Bessell}
  Bessell, M. S \& Brett, J. M. \ 1988, \pasp, 100, 1134
\bibitem[Deharveng et al.(1997)]{Deharveng}
  Deharveng, L., Zavagno, A., Cruz-Gonz\'alez, I., Salas, L., Caplan, J., Carrasco, L. \ 1997, \apj, 317, 459
\bibitem[Elmegreen et al.(1977)]{Elmegreen}
 Elmegreen, B. G., Lada, C. J.  \ 1977, \apj, 214, 725
\bibitem[Feinstein et al.(1986)]{Feinstein}
 Feinstein, A., V\'azquez, R. A., Benvenuto, O. G. \ 1986, \aap, 159, 223
\bibitem[Harris et al.(1976)]{Harris}
  Harris, G. L. H. \ 1976, \apjs, 30, 451
\bibitem[Haisch Lada (2001)]{Haisch}
  Haisch, K. E. \& Lada, E. A. \ 2001, \apj, 553, L153
\bibitem[Hester et al.(2004)]{Hester}
  Hester, J. J., Desch, S. J., Healy, K. R., Leshin, L. A.
  2004, Science, 304, 1116
\bibitem[Hillenbrand et al.(1998)]{Hillenbrand}
  Hillenbrand, L. A., Strom, S. E., Calvet, N., Merrill, K. M., Gatley, I., Makidon, R. B., Meyer, M. R., Skrutskie, M. F. \ 1998, \aj, 116, 1816
\bibitem[Itoh et al.(1996)]{Itoh}
  Itoh, Y., Tamura, M., Gatley, I. \ 1996, \apj, 465, L129
\bibitem[Jones et al.(1981)]{Jones}
  Jones, T. J., Ashley, A., Hyland, A. R., Ruelas-Mayorga, A. \ 1981, \mnras, 197, 413
\bibitem[Meyer et al.(1997)]{Meyer}
  Meyer, M. R., Calvet, N., Hillenbrand, L. A. \ 1997, \aj, 114, 288
\bibitem[Moffat (1972)]{Moffat}
 Moffat, A. F. J. \ 1972, \aaps, 7, 355
\bibitem[Nagashima (1999)]{Nagashima}
	Nagashima, C. et al. \ 1999, in Proc. of Star Formaion 1999, 
	Nagoya, 21-25 June 1999, ed. T. Nakamoto (Nobeyama Radio Observatory), 
	397
\bibitem[Nagayama (2003)]{Nagayama}
	Nagayama, T., et al. \ 2003, \procspie, 4841, 459
\bibitem[Oasa et al.(2006)]{Oasa}  
 	Oasa, Y. et al. \ 2006, \aj, in press
\bibitem[Ogura et al.(2002)]{Ogura}  
  Ogura, K., Sugitani, K., Pickles, A. \ 2002, \aj, 123, 2597
\bibitem[Persson et al.(1998)]{Persson}  
  Persson, S. E., Murphy, D. C., Krzeminski, W., Roth, M., Rieke, M. J. \ 1998, \aj, 116, 2475
\bibitem[Sugitani et al.(1991)]{Sugitani}
  Sugitani, K., Fukui, Y., Ogura, K. \ 1991, \apj, 77, 59
\bibitem[Sugitani et al.(1995)]{Sugitani95}
  Sugitani, K., Tamura, M., Ogura, K. \ 1995, \apj, 455, L39
\bibitem[Tachibana \& Huss (2003)]{Tachibana}
  Tachibana, S., Huss, G. R. \ 2003, \apj, 588, L41
\end{thebibliography}
\end{document}